\documentclass[twocolumn,amsmath,floatfix]{revtex4}
\usepackage{latexsym}
\usepackage{amssymb}
\usepackage{pdfpages}
\usepackage{graphics}

\begin{document}
\newcommand{\ja}{Jakuba\ss a-Amundsen }
\newcommand{\bfx}{\mbox{\boldmath $x$}}
\newcommand{\bfq}{\mbox{\boldmath $q$}}
\newcommand{\bfnabla}{\mbox{\boldmath $\nabla$}}
\newcommand{\bfalpha}{\mbox{\boldmath $\alpha$}}
\newcommand{\bfsigma}{\mbox{\boldmath $\sigma$}}
\newcommand{\bfeps}{\mbox{\boldmath $\epsilon$}}
\newcommand{\bfA}{\mbox{\boldmath $A$}}
\newcommand{\bfP}{\mbox{\boldmath $P$}}
\newcommand{\bfe}{\mbox{\boldmath $e$}}
\newcommand{\bfn}{\mbox{\boldmath $n$}}
\newcommand{\bfW}{{\mbox{\boldmath $W$}_{\!\!rad}}}
\newcommand{\bfM}{\mbox{\boldmath $M$}}
\newcommand{\bfI}{\mbox{\boldmath $I$}}
\newcommand{\bfJ}{\mbox{\boldmath $J$}}
\newcommand{\bfQ}{\mbox{\boldmath $Q$}}
\newcommand{\bfY}{\mbox{\boldmath $Y$}}
\newcommand{\bfp}{\mbox{\boldmath $p$}}
\newcommand{\bfk}{\mbox{\boldmath $k$}}
\newcommand{\bfks}{\mbox{{\scriptsize \boldmath $k$}}}
\newcommand{\bfqs}{\mbox{{\scriptsize \boldmath $q$}}}
\newcommand{\bfxs}{\mbox{{\scriptsize \boldmath $x$}}}
\newcommand{\bfalphas}{\mbox{{\scriptsize \boldmath $\alpha$}}}
\newcommand{\bfs}{\mbox{\boldmath $s$}_0}
\newcommand{\bfv}{\mbox{\boldmath $v$}}
\newcommand{\bfw}{\mbox{\boldmath $w$}}
\newcommand{\bfb}{\mbox{\boldmath $b$}}
\newcommand{\bfxi}{\mbox{\boldmath $\xi$}}
\newcommand{\bfzeta}{\mbox{\boldmath $\zeta$}}
\newcommand{\bfr}{\mbox{\boldmath $r$}}
\newcommand{\bfrs}{\mbox{{\scriptsize \boldmath $r$}}}
\newcommand{\bfps}{\mbox{{\scriptsize \boldmath $p$}}}

\renewcommand{\theequation}{\arabic{section}.\arabic{equation}}
\renewcommand{\thesection}{\arabic{section}}
\renewcommand{\thesubsection}{\arabic{section}.\arabic{subsection}}

\title{\Large\bf Radiative corrections to elastic electron-carbon scattering cross sections in comparison with experiment}

\author{D.~H.~Jakubassa-Amundsen \\
Mathematics Institute, University of Munich, Theresienstrasse 39,\\ 80333 Munich, Germany}


\vspace{1cm}

\begin{abstract}  
The influence of dispersion on the differential scattering cross section in the vicinity of the first diffraction minimum is revisited for collision energies between 200 and 450 MeV.
Transient nuclear excitations in the giant resonance region with angular momentum $L\leq 3$ are taken into consideration within  updated numerics.
Moreover, the deviations of the nonperturbative QED corrections from the conventionally used smooth Born predictions are accounted for.
A qualitative agreement with the measurements is only obtained for the lowest energy, while dispersion within the present model is too small at the higher energies.
\end{abstract}

\maketitle

\section{Introduction}

In order to determine nuclear charge distributions or nuclear charge radii from the measurements of elastic electron scattering \cite{Re82,Ka89,Of91} by means of the comparison with theory \cite{Gu20},
based for spin-zero nuclei on the phase-shift analysis \cite{FW66}, either theory or the experimental data have to be corrected for radiative effects.
The published data are usually corrected for the QED effects by means of subtracting a smooth background,
estimated within the Born approximation. Dispersive effects, due to the transient excitation of the target nucleus during the scattering process, are not at all accounted for.

The visibility of the influence of nuclear excitations on the scattering cross section is expected for collision energies above, say, 30 MeV. 
(For $^{12}$C, nuclear excitations are negligible even at 56 MeV, contributing $< 0.03\%$ at all angles \cite{JR23}.)
They manifest themselves predominantly in the region of the diffraction minima where potential scattering is small.
Light targets, such as $^{12}$C, are favoured since Coulomb distortion is mostly of minor importance such that a description of dispersion within the second-order Born approximation is adequate. 

A comprehensive formulation of this theory is given by Friar and Rosen \cite{FR74}.
In their calculations they use, however, a closure approximation for the sum over the excited states,
and this model, being the only one easily employed,  cannot explain the experimental findings \cite{Gu20,Jaku21}.
  In more advanced dispersion estimates for $^{12}$C the transition to excited states is described in terms of approximate form factors \cite{BC72}.
The above models models are, however,  hampered by only considering the charge transition  form factors. 

The inclusion of the magnetic form factors in addition to the charge form factors, obtained from sophisticated nuclear models, is done in \cite{Jaku22}. This prescription, accounting for nuclear excitations of low angular momentum and excitation energies up to 25 MeV, is used in the present work.

For the QED contributions from vacuum polarization and the vertex plus self-energy (vs) correction, a nonperturbative description is applied. These corrections are represented in terms of potentials which are
added to the nuclear potential in the Dirac equation for the electronic scattering states. For the vacuum polarization the Uehling potential \cite{Ue35} is used, while the vs potential is derived from the respective first-order Born amplitude \cite{Jaku24}.
Within this prescription these QED effects exhibit a structure in the vicinity of the diffraction minima, as is the case for dispersion \cite{Jaku21}, in contrast to their representation in the Born approximation \cite{Ts61}. 
When comparing theory to experiment, this nonmonotonic behaviour has to be taken into account.

The paper is organized as follows. After a short recapitulation of the theory (Section 2), results for the angular distribution of the radiative corrections, including soft bremsstrahlung and dispersion, are presented in Section 3.
The conclusion is given in Section 4. Atomic units ($\hbar=m_e=e=1)$ are used unless indicated otherwise.

\section{Theory}

The differential cross section for elastic scattering from spin-zero nuclei, including the QED corrections and dispersion, is calculated for unpolarized electrons from
$$\frac{d\sigma}{d\Omega_f}\,=\,\frac{|\bfk_f|}{|\bfk_i|\,f_{\rm rec}}\,(1+W_{fi}^{\rm soft}) \,\frac12 \sum_{\sigma_i \sigma_f}$$
\begin{equation}\label{2.1}
\times\, \left[ \,|f_{\rm vac+vs}|^2 + 2\mbox{ Re}\left\{ f_{\rm coul}^\ast ( A_{fi}^{\rm vs(2)} + A_{fi}^{\rm box})\right\}\right],
\end{equation}
where $f_{\rm rec}$ is a recoil factor \cite{Jaku21}, the sum runs over the two polarization states $\sigma_i,\sigma_f$ of the incoming, respectively outgoing electron,
$f_{\rm coul}$ is the scattering amplitude in the Coulombic target potential $V_T$ as obtained from the phase-shift analysis (see, e.g. \cite{YRW} for the special case of $m_e=0$),
and $f_{\rm vac+vs}$ is the scattering amplitude in the potential $V_T+V_{\rm vac}+V_{\rm vs}$ as obtained from the phase-shift analysis.
Here, $V_{\rm vac}$ is the Uehling potential in its parametrized form \cite{FR76} and $V_{\rm vs}$ is the vs potential \cite{Jaku24},
\begin{equation}\label{2.2}
V_{\rm vs}\,=\,-\,\frac{2Z}{\pi} \int_0^\infty d|\bfq|\,\frac{\sin(|\bfq|r)}{|\bfq|r}\,F_L(|\bfq|)\,F_1^{\rm vs}(-q^2),
\end{equation}
generated from the electric vs form factor $F_1^{\rm vs}$ \cite{Va00} (while the magnetic vs form factor $F_2^{\rm vs}$ \cite{BS19} is treated in Born, leading to the amplitude $A_{fi}^{\rm vs(2)}$ in (\ref{2.1}), but this amplitude plays no role at the high collision energies considered here).
$Z$ is the nuclear charge number and $F_L$ is the nuclear charge form factor.

The term $W_{fi}^{\rm soft}$ accounts for the presence of soft bremsstrahlung, cut off by the frequency $\omega_0$ which is determined by the detector resolution at the elastic peak.
Omitting the IR divergence and considering large momentum transfers with $-q^2/c^2 \gg 1$ (where $q=k_i-k_f$ is the difference between initial ($k_i$) and final $(k_f)$ electron 4-momenta),
one has \cite{MT00}
$$W_{fi}^{\rm soft}\,=\,\frac{1}{\pi c} \left\{ \left[ \ln(\frac{-q^2}{c^2})\,-1\right] \ln \frac{\omega_0^2}{E_iE_f}\right.$$
\begin{equation}\label{2.3}
\left. +\,\frac12\,\ln^2(\frac{-q^2}{c^2})\,-\frac12 \,\ln^2 (\frac{E_i}{E_f})\,+\mbox{ Li }(\cos^2 \frac{\vartheta_f}{2})\,-\,\frac{\pi^2}{3}\right\},
\end{equation}
where $E_i$ and $E_f$ are the electronic total energies in the initial, respectively final state,
$\vartheta_f$ is the scattering angle and Li$(x)=-\int_0^x dt\frac{\ln|1-t|}{t}$ is the Spence function.
The fact that bremsstrahlung enters into the cross section by means of the (spin-independent) factor $(1+W_{fi}^{\rm soft})$ guarantees that the spin asymmetry is independent of bremsstrahlung.

Finally, $A_{fi}^{\rm box}$ is the transition amplitude for dispersion, calculated in second-order Born from the Feynman box diagram \cite{FR74,Jaku22},
$$A_{fi}^{\rm box}\,=\,\frac{\sqrt{E_iE_f}}{\pi^2c^3} \sum_{L,\omega_L}\sum_{M=-L}^L \int d\bfp$$
\begin{equation}\label{2.4}
\times \;\sum_{\mu,\nu=0}^3 \frac{1}{(q_2^2+i\epsilon)(q_1^2+i\epsilon)}\,t_{\mu \nu}(p)\,T^{\mu\nu}(LM,\omega_L),
\end{equation}
where the photon propagators have momentum $q_1=k_i-p$ and $q_2=p-k_f$, $p=(E_p/c,\bfp)$ being the momentum of the intermediate electronic state.
The electronic transition matrix element is denoted by $t_{\mu\nu}(p)$, while $T^{\mu\nu}(LM,\omega_L)$ represents the nuclear transition matrix element for the excitation to a  state with energy $\omega_L$, angular momentum $L$ and magnetic projection $M$ and its subsequent decay to the ground state.
In principle, one has to sum  over all possible nuclear excited states, but in the present calculations  restriction is made to the dominant $L\leq 3$ states with energy $\omega_L<25$ MeV.
It turns out that in all cases investigated, the $L=3$ excitations \cite{JR23}  give a negligible contribution to dispersion, and they are omitted henceforth.

\begin{figure}
\vspace{-1.5cm}
\includegraphics[width=11cm]{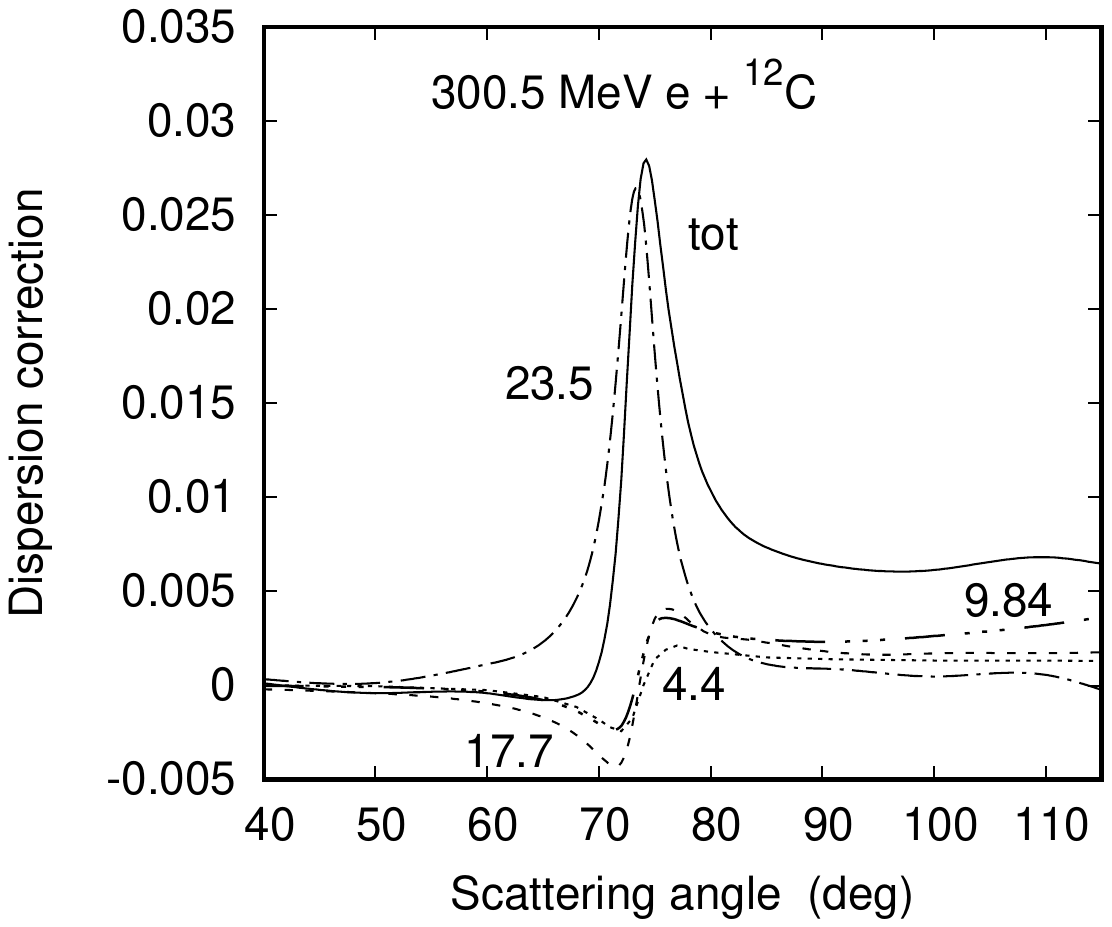}
\vspace{-1.0cm}
\caption
{
Angular distribution of the cross section change $\Delta \sigma^{\rm box}(L,\omega_L)$ by the transient excitation of a state with energy $\omega_L$ and angular momentum $L$ for 300.5 MeV electrons colliding with $^{12}$C.
Shown are the results for the dipole states at 23.5 MeV $(-\cdot -\cdot -)$, at 17.7 MeV $(----)$ and for the quadruploe states at 4.439 MeV $(\cdots\cdots)$ and 9.84 MeV $(-\cdots -)$.
Included is their sum $\Delta \sigma^{\rm box}$ (---------). 
}
\end{figure}
On the other hand, the Born treatment of the QED corrections leads to the following cross section \cite{MT00,Jaku21}
$$\frac{d\sigma^{B-C}}{d\Omega_f}\,=\,\frac{|\bfk_f|}{|\bfk_i|\,f_{\rm rec}}\,\frac12 \sum_{\sigma_i\sigma_f}[ \,|f_{\rm coul}|^2$$
\begin{equation}\label{2.5}
 +\,2\mbox{ Re}\left\{ f_{\rm coul}^\ast (A_{fi}^{\rm vac} +A_{fi}^{\rm vs} +A_{fi}^{\rm box})\right\} +\frac{d\sigma^{\rm soft}}{d\Omega_f}],
\end{equation}
where $A_{fi}^{\rm vs}=A_{fi}^{\rm vs(1)}+A_{fi}^{\rm vs(2)}$ with $A_{fi}^{\rm vs(1)}=F_1^{\rm vs} f_{\rm coul}$,
 $d\sigma^{\rm soft}/d\Omega_f = W_{fi}^{\rm soft} |f_{\rm coul}|^2$, and where the Coulomb distortion is fully included by taking also the amplitudes $A_{fi}^{\rm vac} $ and $A_{fi}^{\rm vs(2)}$ proportional to the Coulombic amplitude $f_{\rm coul}$ as suggested in \cite{Ma69}.

The relative cross section change by the radiative effects is defined by
\begin{equation}\label{2.6}
\Delta \sigma\,=\,\frac{d\sigma/d\Omega_f}{d\sigma_{\rm coul}/d\Omega_f}\,-1
\end{equation}
with $\frac{d\sigma_{\rm coul}}{d\Omega_f}=\frac{|\bfks_f|}{|\bfks_i| f_{\rm rec}} \frac12 \sum_{\sigma_i\sigma_f} |f_{\rm coul}|^2.$

When the product of $W_{fi}^{\rm soft}$ and $A_{fi}^{\rm box}$ in (\ref{2.1}) is neglected (being small of higher order), it is seen that the QED correction $\Delta \sigma^{\rm QED}$ and the dispersive correction $\Delta \sigma^{\rm box}$ are additive.
Therefore one can estimate them separately according to
\begin{equation}\label{2.7a}
 \Delta \sigma^{\rm box} \,=\,\frac{\sum_{\sigma_i\sigma_f} 2 \mbox{ Re }\{f_{\rm coul}^\ast \,A_{fi}^{\rm box}\}}{\sum_{\sigma_i\sigma_f} |f_{\rm coul}|^2}
\end{equation}
$$\Delta \sigma^{\rm QED}\; =
$$
\begin{equation}\label{2.7}
\frac{(1+W_{fi}^{\rm soft})\sum_{\sigma_i\sigma_f}\left[ |f_{\rm vac+vs}|^2+2\,\mbox{Re}\,\{f_{\rm coul}^\ast A_{fi}^{\rm vs(2)}\}\right]}{\sum_{\sigma_i\sigma_f}|f_{\rm coul}|^2}\,-1.
\end{equation}
From the linearity of (\ref{2.7a}) in $A_{fi}^{\rm box}$ it follows that $\Delta \sigma^{\rm box}$ is additive with respect to the contributing excited states,
$\Delta \sigma^{\rm box}=\sum_{L,\omega_L} \Delta \sigma^{\rm box}(L,\omega_L)$.

In the Born approximation for the QED corrections, one has instead of (\ref{2.7})
$$\Delta \sigma^{{\rm QED},B-C}\,=\,$$
\begin{equation}\label{2.8}
\frac{\sum_{\sigma_i\sigma_f}\left[ 2\mbox{ Re } \{f_{\rm coul}^\ast (A_{fi}^{\rm vac} +A_{fi}^{\rm vs})\}+d\sigma^{\rm soft}/d\Omega_f\right]}{\sum_{\sigma_i\sigma_f}|f_{\rm coul}|^2}.
\end{equation}

Assuming that the smooth radiative background, subtracted by the experimentalists, is represented by the Born formula (\ref{2.8}), the correction that has to be applied to theory for the comparison with the experimental data is calculated from
\begin{equation}\label{2.9}
\Delta \sigma^{\rm theor} = \Delta \sigma^{\rm box} \,+\, (\Delta \sigma^{\rm QED} -\Delta \sigma^{{\rm QED},B-C}).
\end{equation}

The experimental radiative corrections are obtained from the measured cross section data relative to the Coulombic phase-shift result \cite{Gu20},
\begin{equation}\label{2.10}
\Delta \sigma^{\rm exp}\,=\,\frac{d\sigma^{\rm exp}/d\Omega_f}{d\sigma_{\rm coul}/d\Omega_f}\,-\,1.
\end{equation}

\begin{figure}
\vspace{-1.5cm}
\includegraphics[width=11cm]{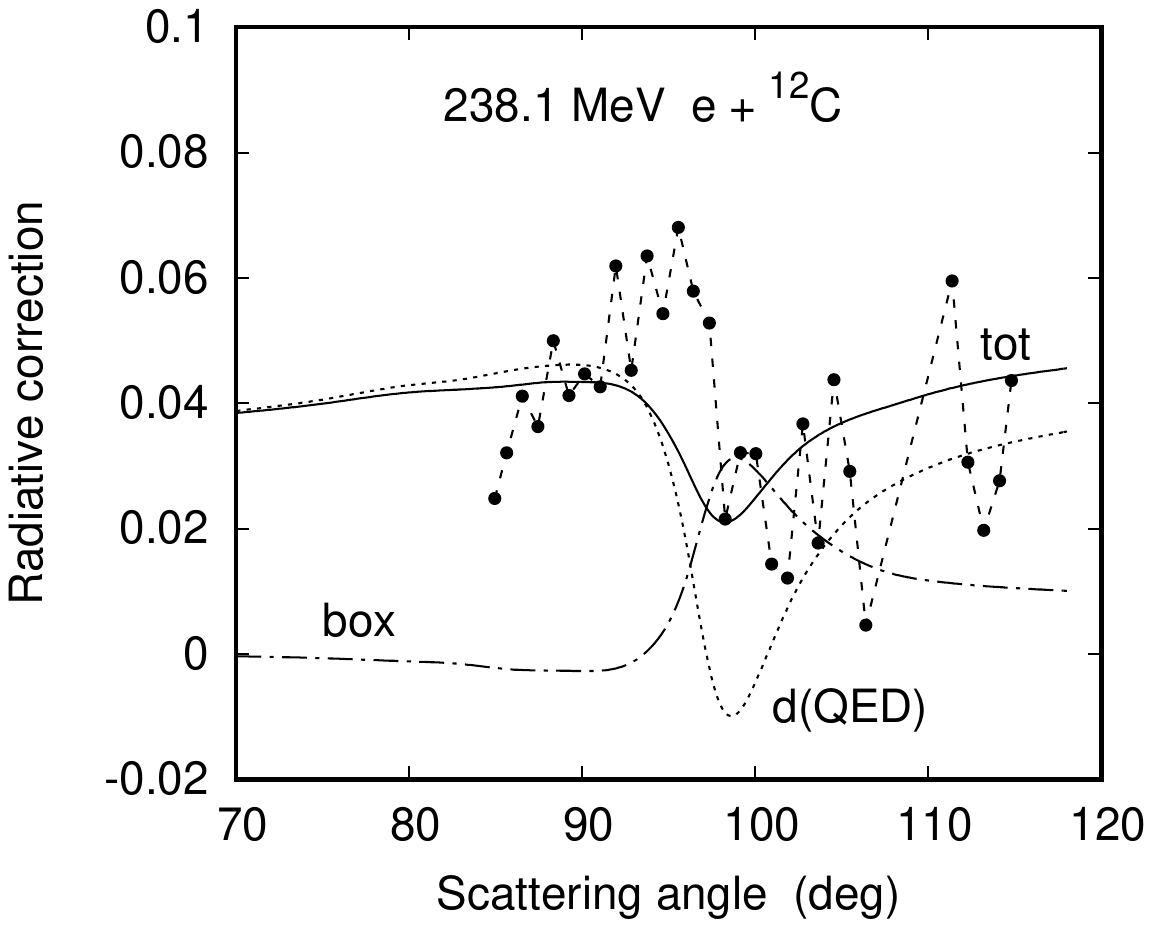}
\vspace{-1.0cm}
\caption
{
Total cross section change $\Delta \sigma^{\rm theor}$ (------) from (\ref{2.9})  in 238.1 MeV $e + ^{12}$C collisions as function of the scattering angle $\vartheta_f$.
Also shown is the result $\Delta \sigma^{\rm box}$ from dispersion $(-\cdot -\cdot -)$, as well as the difference $d(QED)=\Delta \sigma^{\rm QED} -\Delta \sigma^{\rm QED,B-C}\;(\cdots\cdots).$
The experimental points $(\bullet -- \bullet)$, denoting $\Delta \sigma^{\rm exp}$, are deduced from the data of  Offermann et al. \cite{Of91}.
}
\end{figure}
\section{Results for $^{12}$C}

The Coulombic potential $V_T$ is generated from the Fourier-Bessel representation of the nuclear ground-state charge density \cite{deV}.
For obtaining the phase shifts of the electronic scattering states, the Dirac equation is solved with the help of the Fortran code RADIAL by Salvat et al. \cite{Sal}.
The sum over the phase shifts is done with a two-fold convergence acceleration as described in \cite{YRW}. Up to 25000 partial waves are taken into account.
Recoil is -- apart from the prefactor in (\ref{2.1}), $|\bfk_f|/(|\bfk_i| f_{\rm rec})$ -- considered in the phase-shift analysis in terms of a reduced collision energy, $\bar{E} = \sqrt{(E_i-c^2)(E_f-c^2)}$, which is important to match the theoretical diffraction minimum with the experimental one \cite{Jaku21}.

For dispersion, the dominant dipole excitations at 23.5 and 17.7 MeV as well as the quadrupole excitations at 4.439 and 9.84 MeV are taken into account.
The transition densities for the latter state are calculated by Roca-Maza within the Hartree-Fock plus random phase approximation, based on the Skyrme parametrization \cite{Co13}.
The transition densities for the other three states are provided by Ponomarev from the quasiparticle phonon model \cite{Iu12}.
The form factors inherent in $T^{\mu\nu}$ are Fourier-Bessel transforms of these densities \cite{Jaku22}.

Fig.1 compares the angular distribution of the dispersion correction $\Delta \sigma^{\rm box}(L,\omega_L)$ from each of the four considered states with $L\leq 2$ near the first diffraction minimum at $73.8^\circ$ for a collision energy of 300.5 MeV.
It is seen that the dominant contribution is due to the 23.5 MeV dipole state, while the quadrupole states add to $\Delta \sigma^{\rm box}$ mostly  at the larger angles. 
Actually the combined result from the three lower states counterweighs the one from the 23.5 MeV $L=1$ state below the diffraction minimum, but adds to dispersion from
this state at the larger angles. A corresponding behaviour is also found for the other two collision energies considered.

\begin{figure}
\vspace{-1.5cm}
\includegraphics[width=11cm]{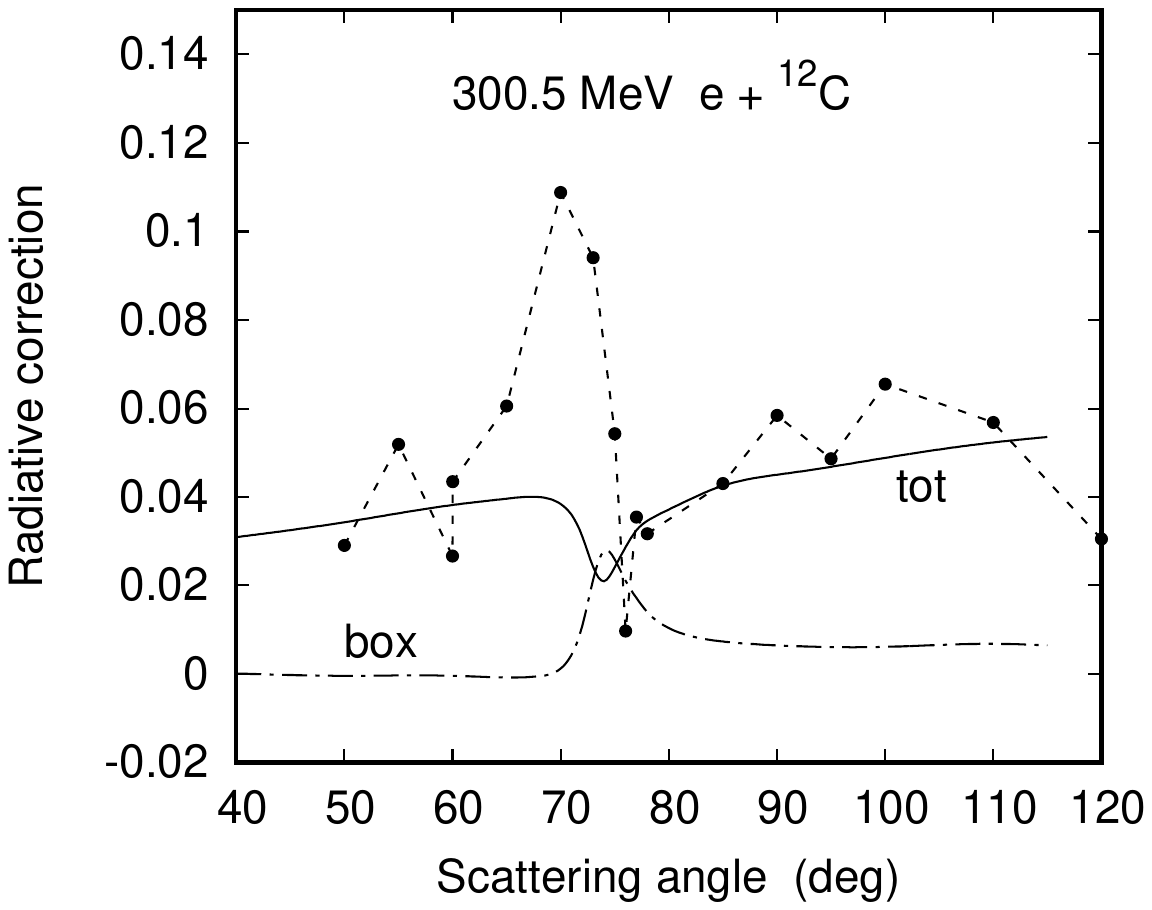}
\vspace{-1.0cm}
\caption
{
Total cross section change $\Delta \sigma^{\rm theor}$ (------)  in 300.5 MeV $e + ^{12}$C collisions as function of the scattering angle $\vartheta_f$.
Also shown is the result $\Delta \sigma^{\rm box}$ from dispersion $(-\cdot -\cdot -)$.
The experimental points $(\bullet -- \bullet)$, denoting $\Delta \sigma^{\rm exp}$, are deduced from the data of Reuter et al. \cite{Re82}.
}
\end{figure}

\begin{figure}
\vspace{-1.5cm}
\includegraphics[width=11cm]{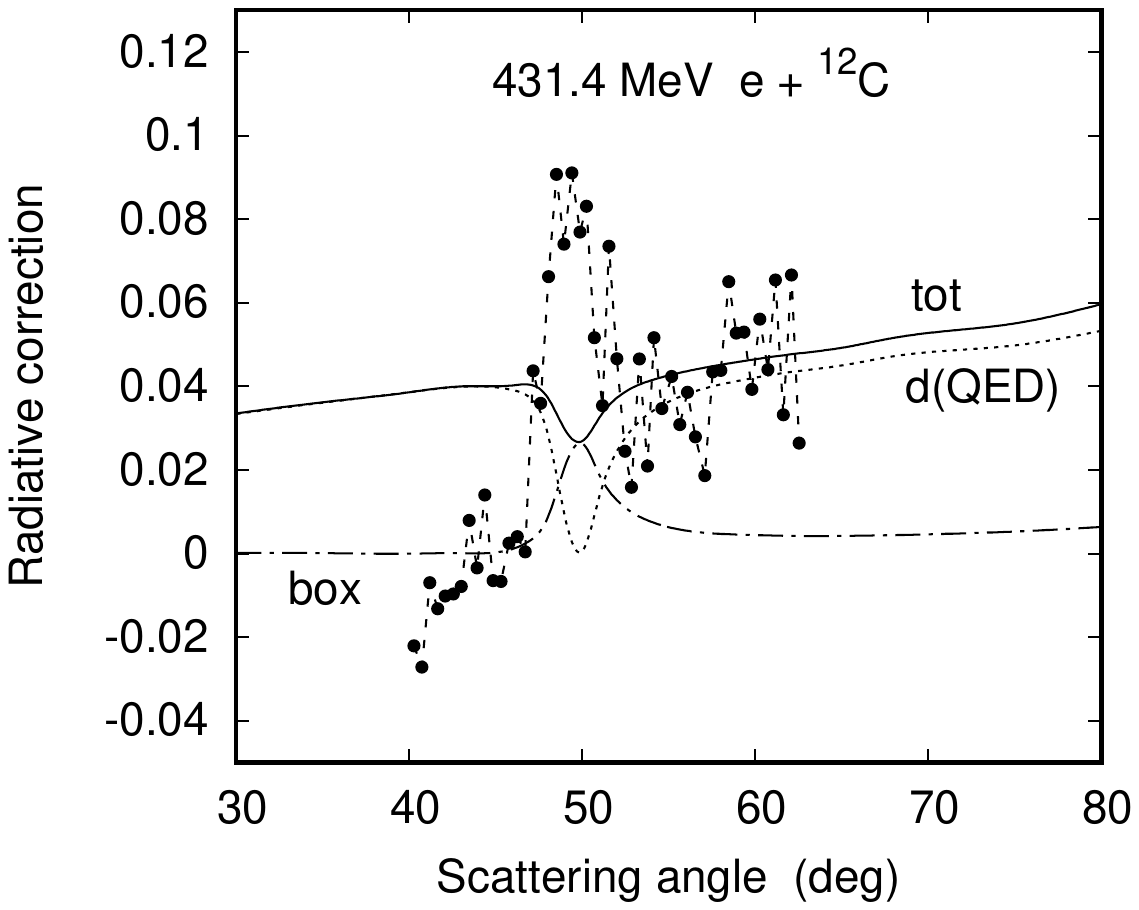}
\vspace{-1.0cm}
\caption
{
Total cross section change $\Delta \sigma^{\rm theor}$ (------)   in 431.4 MeV $e + ^{12}$C collisions as function of the scattering angle $\vartheta_f$.
Also shown is the result $\Delta \sigma^{\rm box}$ from dispersion $(-\cdot -\cdot -)$, as well as the QED difference $d(QED)\;(\cdots\cdots)$, using $\omega_0=0.0863$ MeV for the bremsstrahlung cutoff.
The experimental points $(\bullet -- \bullet)$, denoting $\Delta \sigma^{\rm exp}$, are deduced from the data of Offermann et al. \cite{Of91}.
}
\end{figure}

In Fig.2 the relative cross section change at 238.1 MeV is compared to the one extracted from the experimental data by Offermann et al. \cite{Of91}. 
In contrast to the Friar-Rosen result shown in \cite{Jaku23} (where an average excitation energy of 15 MeV is taken), the present theory for dispersion exhibits a shallow minimum and a pronounced maximum. (Our present results for dispersion differ from those shown in \cite{Jaku22,Jaku23} due to the correction of a spurious enhancement of the form factors, and due to the addition of the  $L=2$ contribution at 9.84 MeV.)
Including the QED correction according to (\ref{2.9}) with $\omega_0 = 0.0476$ MeV (corresponding to the experimental resolution of $2 \times 10^{-4}$), the measured angular distribution is globally quite well reproduced. Only  in the region of the experimental maximum at $95^\circ$ is theory slightly too low.

The situation for 300.5 MeV is displayed in Fig.3.
The experimental data from Reuter et al. \cite{Re82} are in the wings still well described by theory (using $\omega_0=0.0751$ MeV, matching the resolution of $2.5 \times 10^{-4}$). However, our result for dispersion is far too small near the diffraction minimum (being of similar magnitude but of different shape as the Friar-Rosen result shown in \cite{Jaku21}).
A corresponding underprediction is also evident at 431.4 MeV (Fig.4), the highest energy presently considered.

In fact, as seen in Fig.5, the dispersion maximum  decreases with increasing impact energy, in contrast to the experimental data \cite{Gu20}. Such a decrease of dispersion is expected for energies well above the pion production threshold (at 135 MeV) as long as dispersion is solely based on the excitation of the low-lying nuclear states \cite{Go}.

\begin{figure}
\vspace{-1.0cm}
\includegraphics[width=11cm]{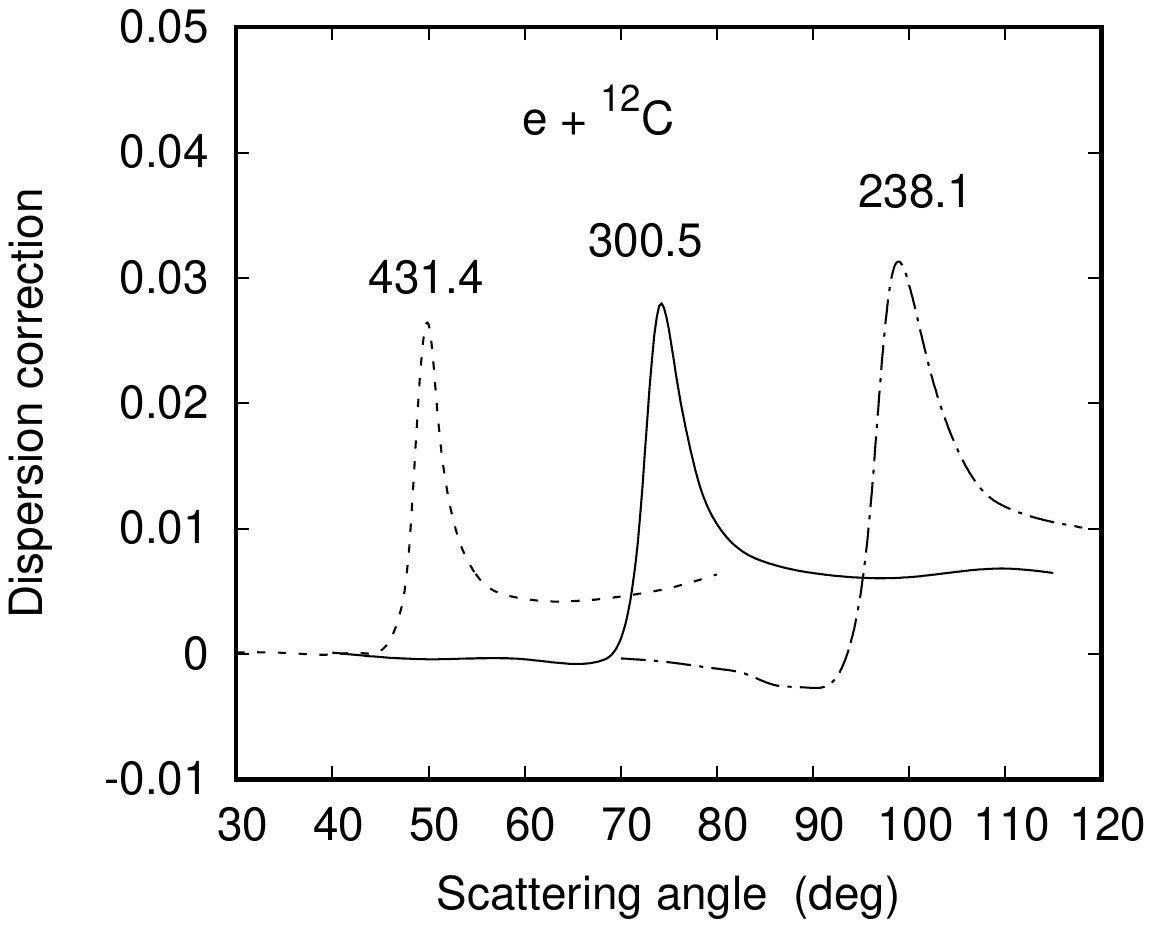}
\vspace{-1.0cm}
\caption
{
Angular distribution of the cross section change $\Delta \sigma^{\rm box}$ by dispersion for electrons colliding with $^{12}$C at energies 238.1 MeV $(-\cdot -\cdot -)$, 300.5 MeV (---------) and 431.4 MeV $(----)$.
}
\end{figure}

\section{Conclusion}

The radiative corrections from dispersion and from the difference between the nonperturbatively treated QED corrections and their Born behaviour were investigated and compared to experiment for the collision energies 238.1, 300.5 and 431.4 MeV.
It was found that in the wings of the first diffraction minimum these QED differences can mostly account for the experimental findings.
Dispersion, on the other hand, is peaked at the position of this minimum.
Hovever, it is severely underestimated at the higher  impact energies.
This indicates that above 200 MeV the consideration of the dominant  nuclear excited states below 25 MeV  is not sufficient to describe dispersion, and hadronic excitations will play an important role. Such excitations are included when dispersion at forward angles  is estimated with the help of the experimental Compton scattering cross section \cite{AM05,GH08}.
In the case of the beam-normal spin asymmetry for high-energy elastic electron scattering, dispersion is for $^{12}$C reasonably well described within such a model \cite{Ko21,An21}.
However, this prescription has not yet been applied to estimate its effect on the differential cross section.
An experimental test of the present theory at energies below the pion production threshold is highly welcome.

\vspace{1cm}



\noindent{\large\bf Acknowledgments}

I would like to thank V.Ponomarev and X.Roca-Maza for calculating the nuclear transition densities.
I would also like to thank M.Gorshteyn for helpful discussions.

\vspace{1cm}

 \end{document}